# Strong Gravitational Lensing as a Probe of Gravity, Dark-Matter and Super-Massive Black Holes

L.V.E. Koopmans (Kapteyn Astronomical Institute)[1],
M. Auger (UCSB), M. Barnabè (Kapteyn/Stanford), A. Bolton (IfA), M. Bradač (UCSB/UCD), L. Ciotti (Bologna), A. Congdon (JPL/Caltech), O. Czoske (Kapteyn), S. Dye (Cardiff), A. Dutton (UCSC), A. Eliasdottir (Princeton), E. Evans (Cambridge), C.D. Fassnacht (UCD) N. Jackson (JBO), C. Keeton (Rutgers), J. Lazio (NRL), P. Marshall (UCSB), M. Meneghetti (Bologna), J. McKean (ASTRON), L. Moustakas (JPL/Caltech), S. Myers (NRAO), C. Nipoti (Bologna), S. Suyu (Bonn), G. van de Ven (IAS), S. Vegetti (Kapteyn), J. Wambsganss (ARI Heidelberg), R. Webster (Melbourne), O. Wucknitz (Bonn), H-S Zhao (St. Andrews)

---

[1] Phone number: +31-(0)50-3636519; Email address: koopmans@astro.rug.nl

# Abstract

Whereas considerable effort has been afforded in understanding the properties of galaxies, a full physical picture, connecting their baryonic and dark-matter content, supermassive black holes, and (metric) theories of gravity, is still ill-defined. Strong gravitational lensing furnishes a powerful method to probe gravity in the central regions of galaxies. It can *(1) provide a unique detection-channel of dark-matter substructure beyond the local galaxy group, (2) constrain dark-matter physics, complementary to direct-detection experiments, as well as metric theories of gravity, (3) probe central supermassive black holes, and (4) provide crucial insight into galaxy formation processes from the dark matter point of view, independently of the nature and state of dark matter.* To seriously address the above questions, a considerable increase in the number of strong gravitational-lens systems is required. In the timeframe 2010-2020, a staged approach with radio (e.g. EVLA, e-MERLIN, LOFAR, SKA phase-I) and optical (e.g. LSST and JDEM) instruments can provide $10^{2-4}$ new lenses, and up to $10^{4-6}$ new lens systems from SKA/LSST/JDEM all-sky surveys around ~2020. Follow-up imaging of (radio) lenses is necessary with moderate ground/space-based optical-IR telescopes and with 30-50m telescopes for spectroscopy (e.g. TMT, GMT, ELT). To answer these fundamental questions through strong gravitational lensing, a strong investment in large radio and optical-IR facilities is therefore critical in the coming decade. In particular, *only* large-scale radio lens surveys (e.g. with SKA) provide the large numbers of high-resolution and high-fidelity images of lenses needed for SMBH and flux-ratio anomaly studies.

## 1. Introduction and Background

The standard cosmological model [1] has been extremely successful in describing the Universe on its largest linear scales. The model is currently based on general relativity (GR) and particle physics (i.e. the other "standard model") and contains a number of ingredients that determine the energy-momentum tensor as function of cosmic time, currently photons, neutrinos, baryons, "dark matter" and "dark-energy". Whereas the first three are well-known and understood, the latter two make up 96% of the energy-density in the Universe and we know very little about them. We do not even know whether gravity behaves as GR predicts on scales far beyond the solar system. Dark matter and energy could simply be "fabrications", confusing modifications of the field equations with modifications of the energy-momentum tensor.

Whereas dark-matter is certainly a testable proposition in direct-detection experiments, dark-energy can thus far only been studied using astrophysics and cosmology. This situation might change, but is not likely foreseen to happen in the next decade. Despite this, even if dark matter and dark energy *are* detected in direct experiments, their properties on large scales, relevant for astrophysics and cosmology, and their phase-space density (e.g. that of dark matter) cannot be extrapolated from the laboratory to the entire Universe. Other complementary experiments are needed to accomplish this.

This has become even more urgent because a number of very puzzling facts have arisen from astrophysical and cosmological observations over the last decades [2], that could point towards some fundamental problems with our understanding of baryons and/or dark matter on nonlinear scales, or even of gravity itself: **(1)** In both spiral and elliptical galaxies with prominent baryonic components, there appears to be a "conspiracy" between dark-matter and baryons, leading to a nearly universal total mass distribution out to the largest measured radii that is very close to "isothermal" (i.e. $\varrho \sim r^{-2}$), with only a small intrinsic scatter between systems. In this category one might also place the tight Fundamental-Plane/Faber-Jackson and Tully-Fisher relations for elliptical and spiral galaxies, respectively, respectively, which seem to show almost no intrinsic scatter. **(2)** In many dark-matter dominated LSB and dwarf galaxies, there appears to be a discrepancy between the observed inner dark-matter density slope and that predicted from numerical simulations. Whereas some galaxies agree with predictions, many others do not, despite the near-universality of this density profile in simulations. **(3)** The abundance of CDM substructures predicted from simulations does not appear to agree with the observed number of dwarf satellites of the Milky Way. Although this can be remedied by assuming that gas in low-mass substructures is removed by radiation and SNa feedback, these purely dark substructures should be present in galaxies in large numbers and remain to be convincingly detected. **(4)** An apparent lower limit of $\sim 10^7$ solar mass on the mass enclosed within a radius 0.3 kpc was recently found for all MW satellites, despite >5 orders of magnitude difference in their luminosity. **(5)** There exists a tight correlation between the central supermassive black hole (SMBH) masses in galaxies and the mass/velocity dispersion of their stellar spheroids.

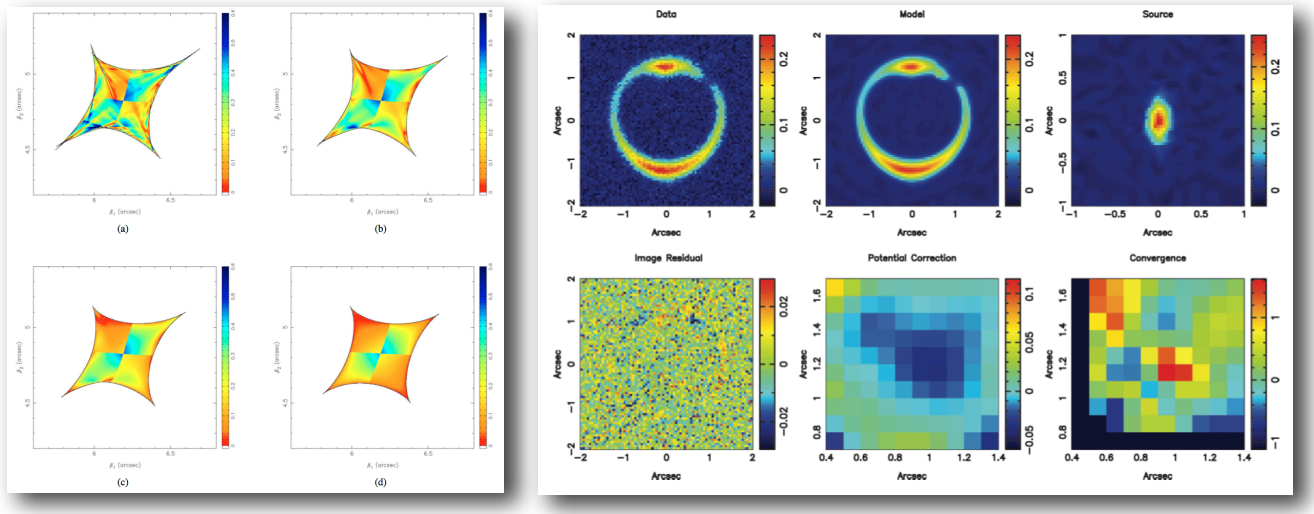

**Fig.1: (left)** Deviations from the cusp relation in an simulated elliptical galaxy. The deviations are due to substructure. Panel a) shows $R_{cusp}$ (see text) for the original mass distribution, whereas panels b)–d) show the cusp relation for the models the substructure is smoothed with a Gaussian kernel with $\sigma_G \sim 1, 2, 5$ kpc for (b), (c) and (d), respectively. (Bradac et al. 2004). **(right)** Results of the linear source and potential reconstruction. The first row shows the original data (left-hand panel), the reconstructed model (middle panel) and the reconstructed source. On the second row, the image residuals (left-hand panel), the potential correction (middle panel) and the substructure convergence (right-hand panel) are shown. The substructure, although weak, is reconstructed at the correct position (Vegetti & Koopmans 2009; [6]).

None of these observations are truly well-understood. For example, dark-matter cores or constant phase-space density and the possible absence of CDM substructure are not predicted and could indicate a modification of either dark-matter properties (e.g. warm DM) or even gravity. Strong gravitational lensing ─ following from metric theories including GR ─ furnishes a powerful technique to address these issues, providing unique constraints on gravity, dark-matter physics and supermassive black holes. *To restrict ourselves, we focus only on galaxy-scale strong lensing*.

## 2. Key Scientific Questions

The key scientific sub-questions that one can extract from the above observational facts are:

    1. *Does CDM substructure exist around galaxies?*

        (i) *What is its dark-matter mass fraction and its mass function (dN/dm)?*

        (ii) *If deviations from CDM predictions are found, what are the implications for dark-matter physics (e.g. its particle properties)?*

    2. *How do SMBH masses depend on galaxy mass and type and how does this relation evolve with cosmic time?*

    3. *Does gravitational lensing agree with standard GR and if not, can other (metric) theories be tested?*

    4. *Does the smooth dark-matter mass distribution in galaxies agree with predictions from the standard cosmological model in its average properties and in its variance?*

        (i) *What is the detailed total and dark matter mass distribution of galaxies as function of their type, mass, stellar population(s) and redshift?*

        (ii) *How do dark matter and baryons (gravitationally) interact in the process of galaxy formation and in their subsequent evolution?*

These questions have certainly been addressed through many different methods, but no satisfactory consensus answer has been obtained yet. We argue that strong lensing can make just that extra step forward in these questions and for several be the only technique to do this in the coming decade. In the next sections we outline the advantages of lensing in probing gravity, dark matter and SMBHs in galaxies and why new radio *and* optical-IR observations en lens surveys are crucial.

## 3. Strong Gravitational Lensing as a Probe of Gravity

Whereas strong gravitational lensing [3] is not the only method to probe the gravitational fields of mass distributions and supermassive black holes, it does provide a clean and unique geometric-optics based method, which does not depend on the nature or dynamical state of the lensing matter, nor requires baryons to be present. A lensed background source is sufficient. Furthermore, lensing effects often maximize at cosmological distances allowing for evolutionary studies.

Below we outline why strong gravitation lensing provides several distinct properties [3] that can address the above key-questions in a unique manner:

- **Surface-brightness conservation** ensures that strong-lens systems exhibiting extended images (i.e. Einstein rings and arcs) are over-constrained and provide information on both the lensed source and the lens mass distribution around the images.

- **The cusp-fold relation** predicts that the sum of image magnifications (parity sign included) tends to zero for a source near the cusp-fold. Deviations from this (i.e. "flux-ratio anomalies") can be used to probe small perturbations in the lens potential on scales smaller than the image separations, e.g. due to CDM substructure.

- **Time delays** between compact lensed images provide complementary information on the lens mass distribution. Through their positions, flux-ratios and time-delays (the latter requires variable sources) the galaxy mass distribution can be probed on all scales.

- **The odd-image theorem** predicts that galaxies have a central image. Absence is thought to be due to the high central stellar density, leading to a large demagnification of the image. In the presence of a SMBH, this theorem breaks down and the central image either disappears, or is replaced by additional images that can be used to measure the mass of the central black hole.

The combination of these geometric-optics properties with deep high-fidelity (S/N≫$10^3$) and high-resolution (≪1″) imaging of strong gravitational lenses can address the posed key-questions in either a unique or complementary manner, that only depends on our understanding of gravity. It is this sole use of metric theories that makes it a very powerful method to study the properties of gravity itself, dark matter and super-massive black holes.

## 4. Direct Detection of Dark-Matter Substructure and SMBHs

Whereas gravitational lensing can address each of the key-questions, we only shortly address two of these where lensing might be the *only* direct way of studying them: CDM substructure and super-massive black holes in galaxies [6,7]. Both are "dark" objects where gravity is the dominant channel of detection and study, without having to assort to complex astrophysics. The other key questions (3 & 4) are deeply connected to these, but will only mentioned in context.

**CDM Substructure:** A solid prediction of the CDM model is the abundant existence of the mass substructure with a mass function very close to $dN/dm \sim m^{-2}$, placing equal amounts of mass in logarithmic bins. A lower-mass cutoff is set by dark-matter physics (e.g. weak interaction, particle velocities during structure formation, etc.) and can be used to measure its properties outside the realm of the laboratory (see also white paper by L. Moustakas). Absence of substructure on scales of ~$10^{8-9}$ solar masses could indicate that our understanding of dark-matter and gravity might have to be reevaluated.

Strong gravitational lensing provides a clean and unique probe of CDM substructure, either through flux-ratio anomalies, time-delay and/or astrometric perturbations, or through direct "gravitational imaging" of their effect on gravitational lensed arcs-rings (e.g. see Fig.1; [7]).

With a modest sample of 10-30 lenses, the technique of "gravitational imaging" can tightly constrain the CDM substructure mass fraction. With a sample of >200 lenses exhibiting lensed arcs-rings (Fig.2), both the nominal CDM substructure mass fraction (0.5%) *and* mass-function slope (1.9) can be recovered to 10% accuracy, assuming a conservative detection threshold of >$10^8$ solar mass [7].

Sample sizes of 10-30 are in the range of existing lens surveys, but samples of ≫100 with high-resolution and high-fidelity extended arcs and Einstein rings are not yet available. Upcoming optical-IR and radio instruments, however, can furnish such numbers in stages over the next decade.

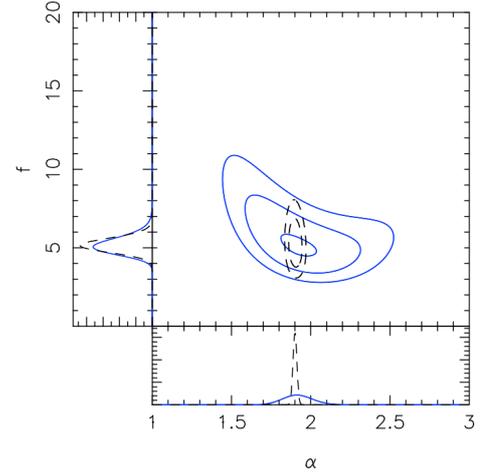

**Fig.2:** Likelihood contours (68, 97, 99.7%) of the recovered CDM mass fraction (f×$10^{-3}$) and slope ($\alpha$) using 200 arc-Einstein-ring systems. Solid contours are without prior on the mass-function slope. Dashed contour include a Gaussian prior around $\alpha=2.0$ with $\sigma=0.1$ (from Vegetti & Koopmans, 2009, in prep. [7]).

Complementary to this, compact lensed radio images could yield a full probability-distribution function of the flux-ratio anomaly parameter $R_{cusp}=\sum \mu_i$ (i.e. the sum of the cusp-fold image magnifications), predicted from simulations (Fig.3), their astrometric perturbations, and through monitoring also anomalous time-delays. *Thus, lensing provides a unique prospect to probe CDM substructure, quantify its properties and probe dark-matter physics on scales beyond the laboratory.*

**Super-Massive Black Holes:** Another very exiting prospect, provided by radio gravitational lenses, is the detection and mass measurement of SMBHs in cosmologically distant galaxies [7]. It provides a direct geometric way of measuring black hole properties as function of cosmological distance, without having to resort to relatively complex astrophysics (e.g. reverberation mapping) or stellar dynamics, both of which are observationally expensive for large samples of SMBHs.

As shown in Fig.3 ([7]), the presence of a SMBH can either "destroy" central images in galaxies, or "create" two additional ones (of opposite parity and offset to one side of the lens center, i.e. easy to recognize and little affected by scattering). The probability for this to happen is small, and outside the dynamic range of current telescopes. However, the large samples of lensed radio sources (≫$10^3$) with large dynamic-range images (e.g. AGN), allows these SMBH "central images" to be detected (see Fig. 3). From this, not only can the presence of SMBHs directly be inferred, but also their masses be measured with considerable accuracy (0.3 dex). Radio facilities with high angular resolution and dynamic-range are required, since faint optical images cannot be detected in the centers of bright lens galaxies. Radio lens systems could build up a sample of systems with SMBH lensed images covering a range in mass and redshift and probe the evolution of the $M_{SMBH}$-$\sigma$ relation without complex modeling.

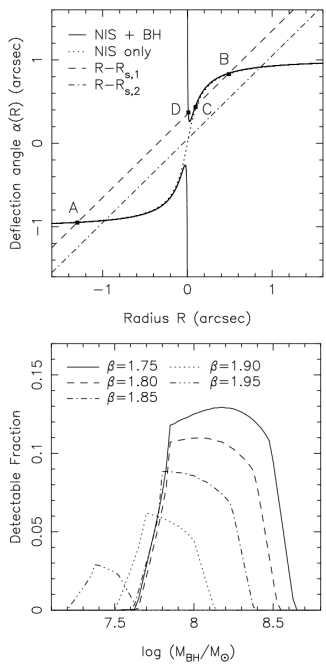

**Fig.3: (Top)** Phenomenology of central images, as illustrated by a cored isothermal model. The deflection profile $\alpha(R)$ is plotted with (solid line) and without (dotted line) a central black hole. Intersections of $\alpha(R)$ and the line R–$R_s$ (where $R_s$ is the source-plane radius) mark the locations of lensed images. The model alone produces three images (A, B, C) of any source within the radial caustic. In the presence of a central black hole, a fourth image (D) is produced for sufficiently misaligned sources ($R_{s,1}$:dashed line), while for well-aligned sources image C is destroyed ($R_{s,2}$: dash-dotted line). **(Bottom)** Detectability of central image pairs produced by a PL galaxy with a supermassive black hole. The fraction of source positions inside the radial caustic is plotted that satisfy image magnification ratio cuts of $\mu_C/\mu_A > 2\times10^{-3}$ and $\mu_D/\mu_C > 10^{-2}$. The lens redshift is 0.5, the source redshift is 1.5, and the Einstein radius is 0.75" for these and all subsequent simulations. We show results for five different profile slopes, spanning the range $1.75 \leq \beta \leq 1.95$. According to the local MBH–$\sigma$ relation, the model galaxy should host a SMBH with log($M_{BH}/M_\odot$)=8.2 ± 0.2. (credit: Rusin, Keeton & Winn 2005; [7]).

*To answer the key-questions, outlined by the two examples discussed above, much larger samples of strong gravitational lenses are required than currently available. These should cover a wide range of galaxies masses, types, environments and redshifts, to minimize statistical uncertainties and biases in the results and allow rare lensing events [4] to be discovered.*

## 5. Large Scale Lens Surveys

**Present Status:** The two largest gravitational lens survey conducted to date are the Cosmic Lens All-Sky Survey (CLASS) and the Sloan Lens ACS (SLACS) Survey (Fig.4; [4]). Candidate selection in the former was done in the radio with the VLA, MERLIN & VLBA (1990-2003), and resulted in 22 lens systems. The SLACS survey (2003-present) combined SDSS spectroscopic preselection with HST imaging follow-up and has yielded ~100 new lens systems. Whereas the ~200 currently known strong lens systems gave a large number of solid scientific results, such as measurements of $H_0$, strong constraints on galaxy density profiles and their evolution, micro-lensing by stars in galaxies, insight in the FP of early-type galaxies, possibly a first indication of CDM substructure through flux-ratio anomalies, etc. [3], many of these studies are limited by the relatively small suitable sample sizes (dozens). *To answer the key-questions posed above and enable trends as function of galaxy type, mass, redshift, etc, to be studied, at least several orders of magnitude more suitable lens systems are required.*

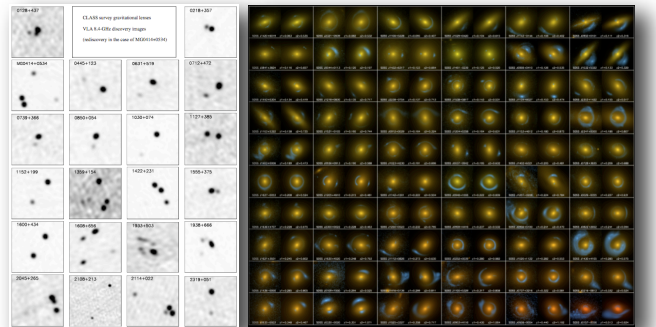

**Fig.4** Strong lenses from the CLASS **(left)** and SLACS **(right)** surveys (see text; [4]), the two largest surveys conducted to date (credit CLASS and SLACS surveys).

**The mid-2010s:** In the coming decade major radio and optical-IR survey telescopes are planned to come online, capable of yielding large samples of galaxy-scale lenses and opening a new discovery space: e.g. in the optical both LSST (>2014) and JDEM (>2015) could yield $10^{4-5}$ new lenses. In the radio the EVLA, e-MERLIN, using e.g. LOFAR for preselection, could yield ~$10^{2-3}$ lenses within the next 5 years, and from the 10% SKA in phase-I (>2016) one expects $10^{3-5}$ lenses as well.

Thus both optical-IR *and* radio facilities are expected to yield $\gg 10^3$ new lenses in the coming decade enabling one to start addressing the key-questions posed above. Until the mid-2010s, however, lens surveys in the optical are likely to dominate because of their superior source number densities on the sky. Whereas some of the key-question can be answered with those optical systems, others (e.g. SMBHs) cannot (see below). We refer to the white paper by P. Marshall for the use of the LSST/JDEM to find lenses in the optical, and below we further outline what large radio-facilities can do.

**Approaching 2020:** As an ultimate objective, ramping up from the use of the e.g. LOFAR, EVLA and e-MERLIN, we focus attention on the SKA (>2016), which can open a whole new realm of strong-lens studies with ~$10^5$ (10% SKA) to ~$10^6$ potential lenses (100% SKA) on every hemisphere. We expect at the end of the next decade, with 100% SKA coming online, that radio lens surveys will gain an edge over optical lens surveys for several reasons: (i) The sky number-density of radio source approaches (or exceeds in deeper integrations) that of optical surveys, (ii) radio facilities can quickly cover the entire sky with multi-beaming, controlled sub-arcsec resolution and milli-arcsec follow-up (VLBI), necessary for detailed substructure and SMBH studies. (iii) Radio facilities, such as SKA, provide spectroscopic redshifts through e.g. HI/CO emission-lines for many lens/source systems, avoiding expensive spectroscopic follow-up. (iv) Lens selection in the radio can be less involved due to the high angular resolution and an advantageous source/lens contrast (Fig.5).

**Radio All-SKA Lens Survey (RASKALS):** To outline the potential rewards of large future radio surveys, we base our the estimates on [5]. One expects 10-100% SKA to detect $10^8$-$10^9$ sources per 20,000 sq. deg. or *$10^5$-$10^6$ potential strong galaxy-scale gravitational lenses*. In this calculation, we assumed (i) an optical depth of $1.4 \times 10^{-3}$ for strong lensing by massive galaxies based on the CLASS survey, (ii) an angular resolution of ~0.25" (SKA baselines ≤180 km at 1.4 GHz), and (iii) a 1.4-GHz continuum survey depth of ~0.5 μJy per pointing (10-σ) (400 MHz bandwidth, 24 hrs of integration, A/T=12,000 $m^2$/K using 75% of 3000 15-m dishes inside 180 km). At this depth, the confusion noise has not been reached and we expect the brighter lenses to be relatively easy to identify (CLASS had a ~100% efficiency), although the efficiency for recognizing and confirming a lens might be lower for fainter sources due to reasons of confusion with complex non-lensed sources. We conservatively set the efficiency at ~10% integrated over all source flux-levels.

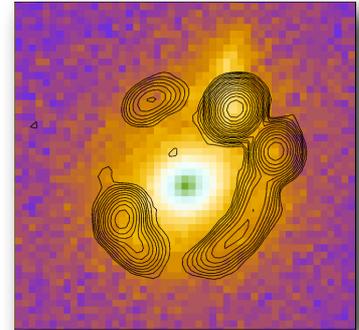

**Fig.5** A CLASS lens observed with the VLA (contours) and in the optical with HST. Note that discovery in the radio reduces confusion between lens and source (credit CLASS survey).

Thus a 10-100% SKA can provide >$10^4$-$10^5$ discoverable strong lens galaxies. [For clusters, the optical depth is ~$2 \times 10^{-5}$, hence ~$10^{3-4}$ strong cluster lenses can be expected in a RASKAL Survey] Given that ~15% of strong lens galaxies are expected to be spiral galaxies, we expect a fair sample of those systems as well. These strong lens systems cover the full range of galaxy types, masses and redshifts, their probability distribution set by their respective integrated lensing cross-sections and magnification biases.

**Optical-IR complementarity:** Complementarity with optical-IR telescopes, however, is crucial for the follow-up of (sub) samples of radio lens systems. Radio and optical data provide different types of information (Fig.5). Joint studies of the dark matter and baryonic content also benefits considerably from optical-IR imaging follow-up, which is particularly important to constrain the stellar/baryonic mass function of galaxies. The redshift range of lenses requires at least 8-10m class telescopes for this. LSST/JDEM are therefore natural partners of radio-facilities. In addition, joint lensing, kinematic and spatially resolved stellar-populations studies can be done to redshifts of z~1 and beyond in the era of 30-50m optical-IR telescopes, using IFU technology. Suitable targets for this, need to be selected from a larger sample of lenses provides by the above radio and optical facilities, since 30-50m telescopes have extremely small fields-of-view.

## 6. New Science Opportunities Through Lensing in 2010-2020

Whereas there are many exciting applications of strong gravitational lensing, several are special. In particular, questions about *CDM substructure and central SMBHs, DM properties and metric gravity theories,* can uniquely be addressed with strong lensing. Beyond the Local Group, gravity is the only way, for example, to seriously study dark substructures and SMBHs.

In this white paper, we aimed to convey that *strong gravitational lensing* provides a gravity-based probe to study "dark" objects through their perturbative effects on strong lensing by galaxies. No other known method can do this in the decade 2010-2020. Whereas dark-matter might decay and produce gamma-ray signals, if these are not detected *and* dark-matter is not found in particle-physics experiments, no other options remain than to find CDM substructures through lensing. Similarly, the measurement of SMBH masses and their properties require detailed dynamical modeling of the few baryons surrounding them. This can easily lead to errors. Gravitational lensing is more direct and circumvents many of these problems. Finally, connected to this, lensing follows directly from metric theories and as such can test them (e.g. GR, TeVeS or other metric theories).

Whereas current lens studies have started to touch upon some of the key-questions, a considerable step forward in the study of CDM substructure and SMBHs, *requires significantly larger samples of strong gravitational-lens systems* with high-resolution and high-fidelity images.

Whereas large ground or space-based optical-IR surveys (e.g. LSST/JDEM) can provide these (see white paper by P. Marshall) for CDM-substructure studies, only radio facilities (e.g. EVLA, e-MERLIN, LOFAR, 10-100% SKA) can provide lens systems with sufficient dynamic range and resolution that lensed images by SMBH can be detected. These radio facilities provide high resolution images and are not affected by poor contrast between lens and lensed source. Looking beyond the next decade, a 10% SKA (>2016), growing into a full SKA, will provide the ultimate "lens engine", potentially discovering >$10^5$ new lens systems.

*We thus conclude that a strong investment in radio and optical-IR facilities, complementing each other, is required in 2010-2020, to (i) discover large samples of strong lens systems, (ii) study CDM substructure and SMBHs and quantify their properties, and (iii) probe dark-matter physics and metric gravity theories, in the decade to come.*

# References

**[1]** Spergel, D.N., et al. 2007.Three-Year Wilkinson Microwave Anisotropy Probe (WMAP) Observations: Implications for Cosmology. Astrophysical Journal Supplement Series 170, 377-408 **[2]** van Albada, T.S., Sancisi, R. 1986. Dark matter in spiral galaxies. Royal Society of London Philosophical Transactions Series A 320, 447-464.; Gerhard, O., Kronawitter, A., Saglia, R.P., Bender, R. 2001. Dynamical Family Properties and Dark Halo Scaling Relations of Giant Elliptical Galaxies. Astronomical Journal 121, 1936-1951; Moore, B., Ghigna, S., Governato, F., Lake, G., Quinn, T., Stadel, J., Tozzi, P. 1999. Dark Matter Substructure within Galactic Halos. Astrophysical Journal 524, L19-L22; Klypin, A., Kravtsov, A.V., Valenzuela, O., Prada, F. 1999. Where Are the Missing Galactic Satellites? Astrophysical Journal 522, 82-92; Swaters, R.A., Madore, B.F., van den Bosch, F.C., Balcells, M. 2003. The Central Mass Distribution in Dwarf and Low Surface Brightness Galaxies. Astrophysical Journal 583, 732-751. Strigari, L.E., Bullock, J.S., Kaplinghat, M., Simon, J.D., Geha, M., Willman, B., Walker, M.G. 2008. A common mass scale for satellite galaxies of the Milky Way. Nature 454, 1096-1097; Magorrian, J., and 11 colleagues 1998. The Demography of Massive Dark Objects in Galaxy Centers. Astronomical Journal 115, 2285-2305. **[3]** Schneider, P., Ehlers, J., Falco, E.E. 1992. Gravitational Lenses. Gravitational Lenses, Springer-Verlag Berlin Heidelberg New; Schneider, P., Kochanek, C.S., Wambsganss, J. 2006. Gravitational Lensing: Strong, Weak and Micro. Saas-Fee Advanced Courses, Volume 33. Springer-Verlag Berlin Heidelberg, 2006 **[4]** Myers, S.T., and 17 colleagues 2003. The Cosmic Lens All-Sky Survey - I. Source selection and observations. Monthly Notices of the Royal Astronomical Society 341, 1-12. Browne, I.W.A., and 21 colleagues 2003. The Cosmic Lens All-Sky Survey - II. Gravitational lens candidate selection and follow-up. Monthly Notices of the Royal Astronomical Society 341, 13-32; Bolton, A.S., Burles, S., Koopmans, L.V.E., Treu, T., Moustakas, L.A. 2006. The Sloan Lens ACS Survey. I. A Large Spectroscopically Selected Sample of Massive Early-Type Lens Galaxies. Astrophysical Journal 638, 703-724. **[5]** Schilizzi, R.T., et al., "Draft Preliminary Specifications for the Square Kilometer Array", v.2.7.1, http://www.skatelescope.org/PDF/Preliminary_SKA_Specifications.pdf; Koopmans, L.V.E., Browne, I.W.A., Jackson, N.J.2004. Strong gravitational lensing with SKA. New Astronomy Review 48, 1085-1094. **[6]** Mao, S., Schneider, P. 1998. Evidence for substructure in lens galaxies? Monthly Notices of the Royal Astronomical Society 295, 587; Dalal, N., Kochanek, C.S. 2002. Direct Detection of Cold Dark Matter Substructure. Astrophysical Journal 572, 25-33; Koopmans, L.V.E. 2005. Gravitational imaging of cold dark matter substructures. Monthly Notices of the Royal Astronomical Society 363, 1136-1144; Vegetti, S., Koopmans, L.V.E. 2009a. Bayesian strong gravitational-lens modelling on adaptive grids: objective detection of mass substructure in Galaxies. Monthly Notices of the Royal Astronomical Society 392, 945-963; Vegetti, S., Koopmans, L.V.E. 2009b Statistics of CDM substructures, in prep.; Bradac, M., Schneider, P., Lombardi, M., Steinmetz, M., Koopmans, L.V.E., Navarro, J.F. 2004. The signature of substructure on gravitational lensing in the LCDM cosmological model. Astronomy and Astrophysics 423, 797-809; Keeton, C.R., Moustakas, L.A. 2008. A New Channel for Detecting Dark Matter Substructure in Galaxies: Gravitational Lens Time Delays. ArXiv e-prints arXiv:0805.0309. **[7]** Mao, S., Witt, H.J., Koopmans, L.V.E. 2001. The influence of central black holes on gravitational lenses. Monthly Notices of the Royal Astronomical Society 323, 301-307; Rusin, D., Keeton, C.R., Winn, J.N. 2005. Measuring Supermassive Black Holes in Distant Galaxies with Central Lensed Images. Astrophysical Journal 627, L93-L96.

*Note that this is a white paper and not intended to be complete in its references. So we opted not to cite all works and ideas that have been described, due to lack of space.*